\documentclass[twocolumn,prb,showpacs,floatfix]{revtex4}

\usepackage{graphicx}
\usepackage{dcolumn}
\usepackage{float}
\usepackage{amsmath}

\newcommand{\comment}[1]{}

\begin{document}


\title{Optical self-energy of superconducting Pb in the THz region}
\author{T. Mori$^1$}

\author{E.J. Nicol$^2$}
\email{nicol@physics.uoguelph.ca}

\author{S. Shiizuka$^{1}$}
\author{K. Kuniyasu$^{3}$}
\author{T. Nojima$^{3}$}
\author{N. Toyota$^{1}$}
\author{J.P. Carbotte$^{4}$}

\affiliation{$^{1}$ Physics Department, Graduate School of Science,
Tohoku University, Sendai 980-8578, Japan}
\affiliation{$^2$Department of Physics, University of Guelph,
Guelph, Ontario N1G 2W1 Canada} 
\affiliation{$^{3}$ Institute for Materials Research and Center for
Low Temperature Science,
Tohoku University, Sendai 980-8577, Japan}
\affiliation{$^4$Department of Physics and Astronomy, McMaster
University, Hamilton, Ontario N1G 2W1 Canada}

\date{\today}

\begin{abstract}
New THz data on the optical conductivity of Pb are presented as well as
a detailed Eliashberg analysis with particular emphasis on phonon-assisted
processes not included in a BCS approach. Consideration of the optical
self-energy instead of the conductivity itself helps highlight the differences
with BCS predictions. Predicted coherence peaks are observed in the optical
scattering rates. Impurities enhance the optical effective mass at zero
frequency by an order of magnitude and induce a large peak at twice
the gap in agreement with theory. This work illustrates the usefulness of
the optical self-energy for the analysis of data.
\end{abstract}

\maketitle

\section{Introduction}

Optical properties of superconducting Pb have been measured in the past
by several groups,
including Palmer and Tinkham\cite{palmer}, who compared their data
with BCS predictions. Joyce and  Richards\cite{joyce} saw phonon
structure in the absorptivity and Farnworth and Timusk\cite{farnworth}
analysed similar data to recover the electron-phonon spectral density
$\alpha^2F(\omega)$. This latter
work was based on approximate equations for the conductivity
derived by Allen\cite{allen} from Fermi golden rule considerations.
Good agreement with the tunneling-derived electron-phonon spectral density
was found\cite{mcmillan}. More elaborate theories of the conductivity
based on the Eliashberg equations and a Kubo formula for the current-current
correlation function were formulated by Nam\cite{nam} and further
elaborated upon in many places\cite{schrieffer} including Scalapino\cite{scalapino} and Marsiglio and Carbotte\cite{marsigliorev,marsiglio}. In this paper
we report new high quality THz measurements on Pb.
In addition to the new data, we 
provide a detailed theoretical analysis 
based on numerical solutions of the Eliashberg equations with  
the known $\alpha^2F(\omega)$~\cite{mcmillan,rowell}. We emphasize particularly
the
phonon-assisted
processes not included in BCS and describe
deviations that these processes introduce.
BCS deals only with the coherent part of the single particle Green's function.
In Pb, the electron-phonon mass renormalization factor $\lambda=1.55$. In
this case, only $1/(1+\lambda)=1/2.55$ 
of the spectral weight resides in the coherent part
and $\lambda/(1+\lambda)=1.55/2.55$  
is found in the incoherent part which describes the 
phonon-assisted processes.

\section{Methods}

The Pb films are deposited onto the (0001) surface of a single-crystalline
sapphire substrate held at room temperature. Four-terminal resistance measurements
on films of 30, 50, and 480 nm in thickness show the superconducting
transitions around 7.2K with the width of $\pm 50$ mK, indicating little
degradation in $T_c$ from the bulk. The present TDTS (Time Domain TeraHertz
Spectroscopy) measurements\cite{Nuss1998} are carried out with the use of a 
commercial spectrometer (RT-20000, Tochigi Nikon Co. Ltd) based on standard
transmission techniques, covering the frequency range of 0.15-2.0 THz. The real
part $\sigma_1(\omega)$ and the imaginary part $\sigma_2(\omega)$ of the
optical conductivity $\sigma(\omega)=\sigma_1(\omega)+i\sigma_2(\omega)$
are directly evaluated by numerically analyzing both the amplitude and
phase of transmitted terawaves through the substrate (0.5 mm in thickness)
with and without the film. The data will be presented for the 50 nm film most
systematically studied.

For a correlated electron system, specifically here a coupled,
electron-phonon system, it has become standard practice to represent
the optical conductivity $\sigma(T,\omega)$  [with $T$, the
temperature,
and $\omega$, the photon energy] in terms of an optical self-energy in
analogy to the quasiparticle self-energy which is introduced to
include
interactions in the one-particle Green's function. By definition,
\begin{equation}
\sigma(\omega)=i\frac{\omega_p^2}{4\pi}\frac{1}{\omega-2\Sigma^{op}(\omega)},
\label{eq:sig2}
\end{equation}
where $\omega_p$ is the plasma frequency. 
The imaginary part of the optical self-energy gives the optical
scattering rate
$1/\tau^{op}(T,\omega)=-2\Sigma_2^{op}(T,\omega)$ 
and its real part gives the optical effective mass ($m^*_{op}(T,\omega)$)
according to 
$\omega[m^*_{op}(T,\omega)/m-1]=-2\Sigma_1^{op}(T,\omega)$.
Further,  $m^*_{op}(T,\omega)/m-1$ is the optical mass renormalization
$\lambda^{op}(T,\omega)$. Here, $m$ is the bare electron mass.
Traditionally, it has been common to display in optical papers
the real and imaginary part of $\sigma(T,\omega)$ but in the recent
literature, it has been recognized that much can be learned from the
scattering rate and effective mass renormalization, particularly for
comparison
with quasiparticle properties. In terms of $\sigma(T,\omega)$ 
\begin{eqnarray}
\frac{1}{\tau^{op}(\omega)} &=&\frac{\omega_p^2}{4\pi}
{\rm Re}\biggl(\frac{1}{\sigma(T,\omega)}\Biggr)=
\frac{\omega_p^2}{4\pi}\frac{\sigma_1}{\sigma_1^2+\sigma_2^2} ,\label{eq:tau}\\
\omega\frac{m^*_{op}(T,\omega)}{m} &=&-\frac{\omega_p^2}{4\pi}
{\rm Im}\Biggl(\frac{1}{\sigma(T,\omega)}\Biggr) =
\frac{\omega_p^2}{4\pi}\frac{\sigma_2}{\sigma_1^2+\sigma_2^2} .
\label{eq:lam}
\end{eqnarray}

The Drude conductivity is recovered when all interactions  are
neglected
and only impurity  scattering is included through a frequency and
temperature independent scattering rate $1/\tau^{imp}$. In this case,
the real part of $\Sigma^{op}(T,\omega)$ which must be related to
its imaginary part by Kramers-Kronig is zero. For coupling
to a single oscillator of energy $\omega_E$  in the normal state
at zero temperature\cite{CSH}
\begin{equation}
\frac{1}{\tau^{op}(T=0,\omega)}=\frac{2\pi A}{\omega}
(\omega-\omega_E)\theta(\omega-\omega_E) ,
\label{eq:taueinstein}
\end{equation}
where $A$  is the coupling between electrons and phonons and
\begin{equation}
\lambda^{op}(\omega)=-\frac{2A}{\omega}\biggl[\ln\biggl|\frac{\omega_E+\omega}{\omega_E-\omega}\biggr|+\frac{\omega_E}{\omega}\ln\biggl|\frac{\omega_E^2-\omega^2}{\omega_E^2}\biggr|\biggr] .
\end{equation}
For $\omega\to 0$, $\lambda^{op}(\omega)=2A/\omega_E\equiv\lambda$, the
quasiparticle electron-phonon mass renormalization. In this model,
$\sigma(\omega)$
has two parts. One part is
 a delta function at $\omega=0$ coming from the region
$\omega<\omega_E$ in (\ref{eq:taueinstein}) where there is no scattering. If
impurity scattering is included, the $\delta$-function broadens into a
Lorentzian
\begin{equation}
\sigma_L(\omega)=i\frac{\omega_p^2}{4\pi}\frac{1}{\omega(m^*_{op}/m)
+i/\tau^{imp}}.\label{eq:siglor}
\end{equation}
The second part which is finite
only for $\omega>\omega_E$ is a boson-assisted part:
\begin{equation}
\sigma_B(\omega)=i\frac{\omega_p^2}{4\pi}\biggl[\omega\biggl(\frac{m^*_{op}(\omega)}{m}\biggr)
+i\biggl(\frac{1}{\tau^{imp}}+\frac{2\pi A}{\omega}(\omega-\omega_E)\biggr)\biggr]^{-1} .\label{eq:sigbos}
\end{equation}
Further, if we approximate in (\ref{eq:siglor}) $m^*_{op}(\omega)/m$
by
$1+\lambda$, its zero frequency limit which is exact for small
$\omega$,
we obtain a coherent Drude part
\begin{equation}
\sigma_D(\omega)=i\frac{\omega_p^2}{4\pi}\frac{1}{\omega(1+\lambda)
+i/\tau^{imp}} ,\label{eq:sigdru}
\end{equation}
with optical spectral weight of $\omega_p^2/[8(1+\lambda)]$
and effective scattering rate of $1/[\tau^{imp}(1+\lambda)]$. The
remaining
spectral weight is to be found in the incoherent piece
(\ref{eq:sigbos})
and is $\omega^2_p\lambda/[8(1+\lambda)]$ which follows on application
of the optical sum rule
\begin{equation}
\int_{0}^\infty\sigma_1(\omega)d\omega=\frac{\omega_p^2}{8}.
\end{equation}

\section{Results}

\begin{figure}[t]
  \centerline{\includegraphics[width=4.30in]{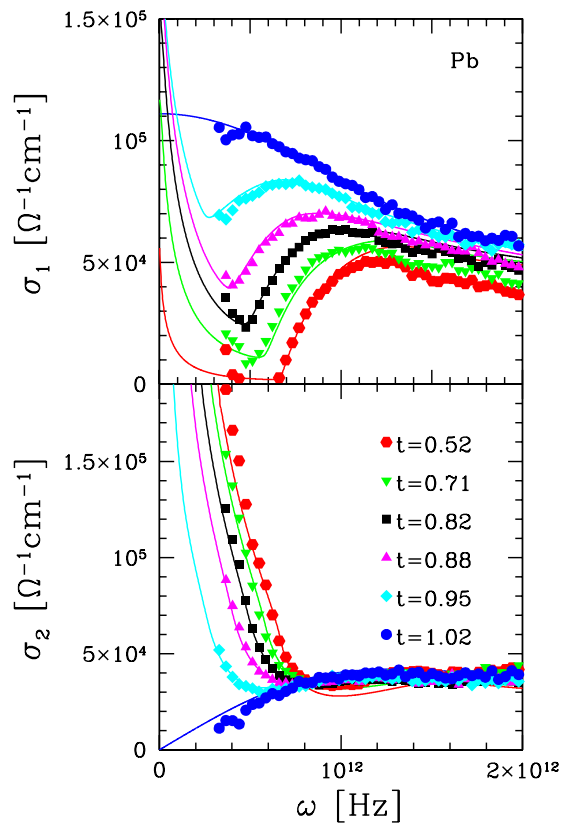}}%
\caption{(color online) Real and imaginary part of the optical
conductivity, $\sigma_1(T,\omega)$ and $\sigma_2(T,\omega)$,
respectively, versus $\omega$ at various reduced temperatures $t=T/T_c$. 
The points
are the data and the lines are theory. 
}
\label{fig1}
\end{figure}
In Fig.~\ref{fig1}, we present our experimental (points) and
theoretical (lines) results for 6 temperatures as labelled 
by reduced temperature $t=T/T_c$. In
the top frame, we show the real part of the conductivity
$\sigma_1(\omega)$
in $\Omega^{-1}\rm{cm}^{-1}$ versus $\omega$ up to a frequency
of $2\times 10^{12}$ Hz while in the lower frame
we show the imaginary part $\sigma_2(\omega)$. 
For the real part, we note a kink in both data and theory at an intermediate
frequency which decreases in frequency, 
with increasing temperature. This arises because there are 
two separate absorption processes possible. At the lowest frequencies there can
be direct absorption from the thermally excited quasiparticles. This contribution
provides a Drude-like peak to $\sigma_1(\omega)$. At frequencies above
twice the gap, direct absorption through the creation of two excitations out
of the condenstate becomes possible. In addition, there is another incoherent
phonon-assisted contribution to the absorption and this is the part of
most interest in this work. However, it cannot easily be
separated out of the real part of the conductivity. We will see that 
consideration of the optical self-energy $\Sigma^{op}$ defined in 
Eq.~(\ref{eq:sig2}) helps in this regard. First, however, we note that the fit
between theory and the data is very good at low frequency in all cases.
We have used the data in the normal state ($t=1.02$) to determine
the elastic scattering rate $1/\tau^{imp}$ and the plasma frequency
$\omega_p$.
Turning next to the imaginary part of $\sigma$ shown in the lower frame of 
Fig.~\ref{fig1}, we note the same level of agreement and 
the points for $t=1.02$
which are in the normal state also show good agreement with the theory.
In general, the superconducting properties of
Pb, while very different from those predicted by BCS, have been remarkably
well-described by Eliashberg theory\cite{carbotte} over a broad range
of experiments
and our new data shown here poses no exception. This is important given that
in the past, it is has been argued that vertex corrections and other
aspects of electromagnetic theory were needed to understand the data in
Pb and yet this appears not to be the case here. With this level of agreement
between theory and data, we should now be able to proceed to test for more
subtle effects due to the phonon-assisted processes in the absorption.

\begin{figure}[t]
  \centerline{\includegraphics[width=4.30 in]{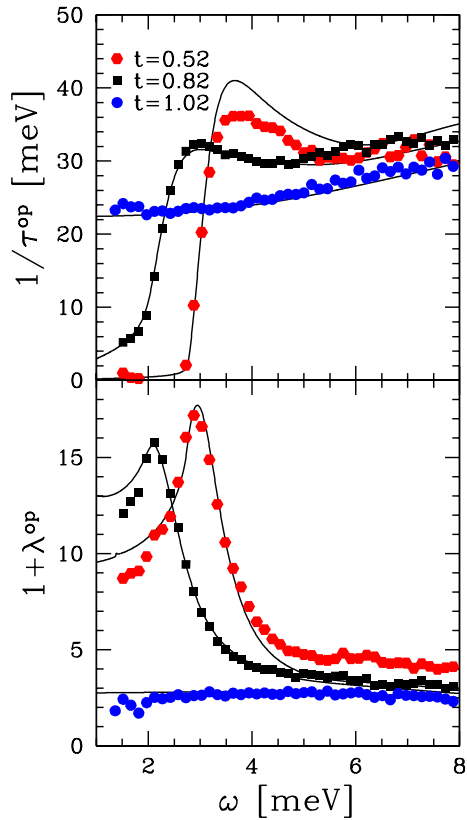}}%
\caption{(color online) The optical scattering rate $1/\tau^{op}(T,\omega)$
and mass renormalization $[1+\lambda^{op}(T,\omega)]$ vs $\omega$.
The points are experiment and the lines are theory. The data for
$t=1.02$ are in the normal state just above $T_c$ and the other two
data sets are below $T_c$ in the superconducting state.
}
  \label{fig2}
\end{figure}

\begin{figure}[t]
  \centerline{\includegraphics[width=4.30 in]{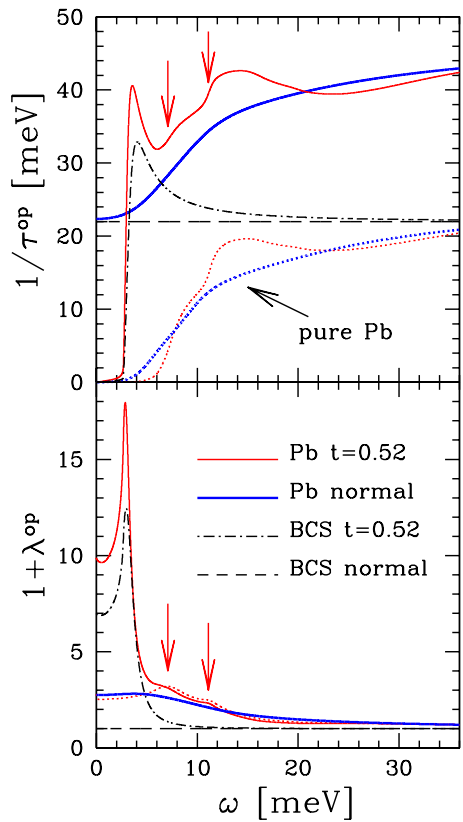}}%
\caption{(color online) The calculated optical scattering rate
$1/\tau^{op}(T,\omega)$
and optical mass renormalization $[1+\lambda^{op}(T,\omega)]$ for a large
energy range.
Results are for our Pb calculation with
impurity scattering rate of 22 meV. Red (thin line)
is in the superconducting state 
and blue (thick line) 
is for the normal state, both at $t=0.52$. We compare with BCS
(dot-dashed and dashed black curves) and pure Pb (dotted curves). The vertical arrows indicate
the peaks in $\alpha^2F(\omega)$ shifted by $2\Delta$.
}
  \label{fig3}
\end{figure}

To see separately, more clearly, the role of the coherent absorption
processes included in BCS theory and described by the coherent
part of the electron Green's function $G$, and the incoherent
phonon-assisted processes described by the incoherent part of $G$,
it is convenient to deal with the real and imaginary parts of the
optical self-energy defined in Eq.~(\ref{eq:sig2}). Equations~(\ref{eq:tau})
and (\ref{eq:lam}) are for the optical scattering rate and effective mass,
respectively. These are shown in Fig.~\ref{fig2}. 
The top frame gives 
$1/\tau^{op}(\omega)$ and the bottom $[1+\lambda^{op}(\omega)]$
at three temperatures. To agree with the experimental data it was
necessary to include an impurity scattering rate of 22 meV. This is 
the only scattering that would be included in a BCS theory for
which the normal state conductivity would be a pure Drude form. The 
optical scattering rate corresponding to this Drude is a constant
flat line independent of frequency and temperature. We see clearly in
the curve for the normal state that significant deviations from
this constant occur. These are due to the boson-assisted processes which
become even more important as the frequency is increased beyond the
range of the data shown. This is shown in the top frame of Fig.~\ref{fig3},
where we present our theoretical results over an extended frequency range
up to 36 meV, well beyond the end of the phonon spectrum. For clarity,
only the $t=0.52$ (superconducting state) and $t=1.02$ (normal state)
curves are shown.
 Starting with the normal state (thick blue curve), we note a large deviation from
the horizontal dashed black line at 22 meV for
elastic impurity scattering. The additional
frequency-dependent scattering, due to boson-assisted processes, adds on
to $1/\tau^{imp}$ according to Eq.~(\ref{eq:taueinstein}) (in the case
of a delta function spectral density). For $\omega\to\infty$ this contribution
will saturate at $2\pi A$ (see Eq.~(\ref{eq:taueinstein})) even for a
distributed $\alpha^2F(\omega)$ spectrum.

Turning next to the superconducting state (thin red curve, top frame),
we see larger boson structures with the vertical arrows indicating the
transverse and longitudinal peaks in the phonon frequency distribution
$F(\omega)$ of Pb shifted by $2\Delta$. These are preceded by a coherence
peak in the region above $2\Delta$ which reflects the increased density of
final states for scattering in the superconducting as compared with 
normal state. This curve is strikingly different from the dot-dashed
 black curve
which would apply to a BCS superconductor. 
An interesting
comparison is offered in the dotted curves which were also calculated
with the Pb $\alpha^2 F(\omega)$ spectrum, but without residual scattering,
i.e. $1/\tau^{imp}=0$. An important point to note is that no coherence peak
is seen in the dotted thin red curve because this is an impurity effect, but
at higher frequencies the dotted curve is very close to the solid
one once it is displaced upward by 22 meV. This holds exactly for the 
normal state data, blue dotted (pure) and solid (with residual scattering).
Returning to the lower frame of Fig.~\ref{fig2}, we show the optical
effective mass renormalization $[1+\lambda^{op}(T,\omega)]$.
The solid blue dots apply to the normal state at $t=1.02$ just above $T_c$.
Note that as $\omega\to 0$ $\lambda^{op}\simeq 1.7$ which is just
slightly larger than the zero temperature quasiparticle
mass renormalization for Pb, which is 1.55. This is expected
since it is known that the mass renormalization increases slightly
with increasing temperature. Furthermore, the normal state
effective mass has only a small frequency dependence predicted and 
this is observed.
There is no dependence on impurity scattering as it drops out of this
curve. (This fact can be understood simply. The Kramers-Kronig (KK) 
transform of a constant is zero so that
a constant scattering rate corresponds to no mass renormalization.) This is in
striking contrast to the superconducting state for which impurities
profoundly alter the optical curves. This is seen in the other two curves
for $t=0.52$ (red hexagons) and $t=0.82$ (black squares). Note first the very
large peaks which occur at twice the gap edge $2\Delta$. These peaks
have not been emphasized before. They would not be there in the clean
case (no impurities). Their origin is easily understood from our
results for the scattering rate shown in the top frame. We see the sharp,
almost vertical, rise
in $1/\tau^{op}(\omega)$  at $2\Delta$ with the energy scale of the rise
given by the value of $1/\tau^{imp}=22$ meV. The KK transform of a step at $\omega_E$ is
a logarithm with a singularity at $\omega_E$. The larger the step, the larger the log
singularity. Of course, here we do not quite have a vertical step and hence the
corresponding peak in the effective mass is smeared. Another 
feature to note is the $\omega\to 0$ limit of $\lambda^{op}$.
We note that  Eq.~(\ref{eq:lam}) can be
rewritten as
\begin{equation}
\frac{m^*_{op}(T,\omega)}{m} =
\frac{\omega_p^2}{4\pi}\frac{\omega\sigma_2}{(\omega\sigma_1)^2+(\omega\sigma_2)^2}
\label{eq:lam2}
\end{equation}
and  that in the superconducting state in the limit of $\omega\to 0$
we get
\begin{equation}
\frac{m^*_{op}(\omega=0)}{m} =
\frac{\omega_p^2}{4\pi}\lim_{\omega\to
  0}\frac{1}{\omega\sigma_2}=\frac{n}{n_s},
\end{equation}
where $n$ is the electron density in the normal state
and $n_s(T)$ is the superfluid density at temperature $T$.
For $T=0$, direct strong-coupling calculations of $n_s(T=0)$ show
that it is very nearly equal to $n/(1+\lambda)$ which leads to 
$m^*_{op}(T=0,\omega=0)/m\simeq1+\lambda$,
same as for the normal state, and reflects the fact that only the 
coherent part of the optical conductivity participates in the
superfluid
condensation. Returning to Eq.~(\ref{eq:lam2}), one can verify that
in the normal state, substitution of the Drude form (\ref{eq:sigdru})
again gives $m^*_{op}/m=1+\lambda$
as we expect.  If impurities are included, as we have done in our work,
$n_s$ is no longer $n/(1+\lambda)$ but in the dirty limit
has instead the approximate form (in renormalized BCS)\cite{marsigliorev}
\begin{equation}
n_s\simeq\frac{n}{1+\lambda}\biggl[\frac{\pi}{2\alpha}
-\frac{1}{2\alpha\sqrt{\alpha^2-1}}
\ln\biggl|\frac{\alpha+\sqrt{\alpha^2-1}}{\alpha-\sqrt{\alpha^2-1}}
\biggr|\biggr]
\end{equation}
for $\alpha>1$ where $\alpha=1/[2\Delta\tau^{imp}(1+\lambda)]$,
which gives $1+\lambda^{op}(T=0,\omega=0)=8.3$.
The exact numerical result (Fig.~\ref{fig2}, bottom frame) is 9.5
at $T/T_c=0.52$. Most of the difference can be assigned to the small
reduction in $n_s(T)$ between $T=0$ and $t=T/T_c=0.52$.

In the lower frame of Fig.~\ref{fig3}, we show additional theoretical
results on the optical effective mass which range over a larger frequency interval
beyond that covered by the THz data. The dot-dashed
 black curve is the BCS result
(for $1/\tau^{imp}=22$ meV) while the solid thin red curve includes as well
the incoherent phonon-assisted Holstein processes. We see clearly in
this curve an image of the Pb phonon frequency distribution with transverse
and longitudinal phonon peaks shifted by $2\Delta$ indicated by vertical arrows.
The thin red dotted lines were obtained for pure Pb (i.e. $1/\tau^{imp}$ is set
equal to zero and the only source of scattering comes from
the inelastic part described by $\alpha^2F(\omega)$). In this
case there is no peak in $[1+\lambda^{op}]$ at twice the gap but at
higher frequencies the dotted and solid curves agree. For the normal
state at $t=1.02$, the solid thick blue curve applies. 
Note
that for $\omega$ larger than a few times the phonon energy, normal and
superconducting curves merge and all tend toward 1, i.e. $\lambda^{op}(\omega)
\to 0$ as $\omega\to \infty$. For BCS in the normal state, $\lambda^{op}=0$
for all $\omega$ as shown by the long-dashed curve.

\section{A Sum Rule}

\begin{figure}[t]
  \vspace*{-2.8 cm}%
  \centerline{\includegraphics[width=4.30 in]{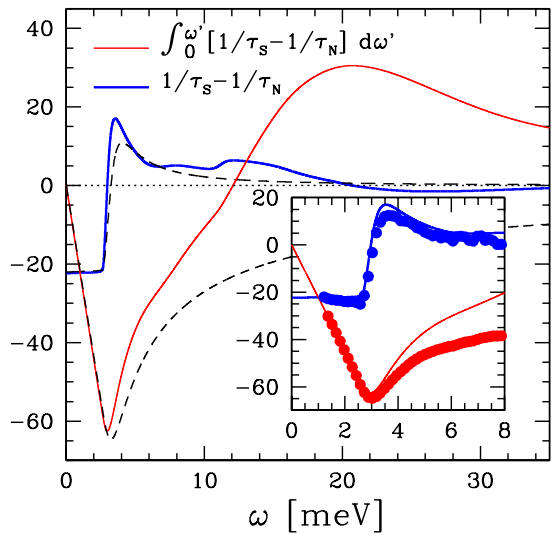}}%
\caption{(color online) The difference
$(1/\tau^{op}_S-1/\tau^{op}_N)$ vs $\omega$ shown up to 35 meV 
and the partial sum, defined as $I(\omega)$ in the text. The
black dashed curves are for the BCS case.
In the inset, we compare data with theory.
}
  \label{fig4}
\end{figure}

Recently, there have
been several discussions of sum rules on $1/\tau^{op}$ in the 
literature\cite{marsiglio02,chubukov,abanov}. In particular,
it was noted that there exists a differential sum rule on the difference
in optical scattering rate between the superconducting and normal state.
Defining the partial sum $I(\omega)$ as 
\begin{equation}
I(\omega)=\int_0^\omega\biggl[\frac{1}{\tau^{op}_S(\omega^\prime)}-\frac{1}
{\tau^{op}_N(\omega^\prime)}\biggr]d\omega^\prime
\label{eq:I}
\end{equation}
it can be shown\cite{chubukov} that $I(\Omega\to\infty)=0$
for any value of $1/\tau^{imp}$ and $\Delta$. The energy scale for arriving
at the asymptotic limit is set by $\Delta$ even in the dirty limit. As defined,
in BCS\cite{chubukov} $I(\Omega)$ is negative and has its maximum absolute
value at $2\Delta$ after which it begins to decrease towards zero, but always
remaining negative. Our results for the case of Pb are shown in  Fig.~\ref{fig4}
and deviate significantly from the BCS behaviour just described.
The solid thick blue curve gives the result for $1/\tau^{op}_S(\omega)-
1/\tau^{op}_N(\omega)$ at $t=0.52$ and shows a rapid increase
through 0 at twice the gap, after which a coherence peak is seen followed
by boson structure beyond which the curve crosses the axis again. 
The solid thin red curve is the result for the partial sum
$I(\omega)$. The dashed black curves are the BCS results for comparison.
The crossing of the horizontal axis at a frequency slightly above
the end of the phonon spectrum in the case for Pb (thin red curve) is due to the
inelastic scattering processes and is not seen in the BCS curve. Note also the large positive
peak.

\section{Conclusions}

We have presented new THz data on the optical conductivity of Pb and compared 
them with Eliashberg results with an emphasis on phonon-assisted (Holstein)
processes not part of BCS. We find that these show up most clearly
in the optical self-energy rather than in the conductivity itself.
In particular, the impurity component of the total scattering is seen 
directly in the optial scattering rate as a large constant which can easily be
subtracted out to reveal directly the phonon-assisted part. In the normal
state impurity scattering drops out of the optical effective mass
leaving only phonon renormalizations which show excellent agreement between
theory and experiment. By contrast, in the superconducting state, the
impurities provide a large renormalization of the zero frequency effective mass and cause a peak to appear at twice the gap in the optical
mass $[1+\lambda^{op}(T,\omega)]$ which we predict here and observe
in the experiment. 

\begin{acknowledgments}
We thank S. Goshima, M. Mori, T. Suzuki and H. Matsui for
valuable assistance and advice in TDTS experiments. The experimental
work at Sendai has been supported by Grants-in-Aid for Scientific Research
(09440141 and 15201019) from the Ministry of Education, Culture, Sports,
Science and Technology, Japan.
This work has also been supported by the Natural Science and
Engineering Council of Canada (NSERC) and the Canadian Institute
for Advanced Research (CIFAR).
\end{acknowledgments}

\end{document}